\documentclass{ws-ijmpa}

\begin{document}

\markboth{Parganlija, Giacosa, Rischke, Kovacs and Wolf}
{A Linear Sigma Model with Three Flavors and Vector and Axial-Vector Mesons}

%
\catchline{}{}{}{}{}
%

\title{A LINEAR SIGMA MODEL WITH THREE FLAVORS AND VECTOR AND AXIAL-VECTOR MESONS\footnote{Prepared for the Proceedings of the 11th International Workshop on Meson Production, Properties and Interaction (MESON 2010), in Krakow, Poland, 10-15 June 2010.}
}

\author{DENIS PARGANLIJA$^{\text{(a)}}$, FRANCESCO GIACOSA$^{\text{(a)}}$, DIRK H. RISCHKE$^{\text{(a,b)}}$ 
}

\address{$^{\text{(a)}}$Institute for Theoretical Physics, Johann Wolfgang Goethe University, Max-von-Laue-Str. 1\\
D--60438 Frankfurt am Main, Germany and \\$^{\text{(b)}}$Frankfurt Institute for Advanced Studies, Ruth-Moufang-Str. 1 \\D--60438 Frankfurt am Main, Germany 
\\
parganlija@th.physik.uni-frankfurt.de}

\author{P\'{E}TER KOV\'{A}CS AND GY\H{O}RGY WOLF}

\address{Research Institute for Particle and Nuclear Physics\\
H-1525 Budapest, POB 49, Hungary\\
wolf@rmki.kfki.hu}

\maketitle


\begin{abstract}
We outline the extension of the globally chirally invariant $N_f = 2$ linear sigma model with vector and axial-vector degrees of freedom to $N_f=3$. We present preliminary results concerning the scalar meson masses.

\end{abstract}

\ccode{PACS numbers: 12.39.Fe, 12.40.Yx}

\section{Introduction}	

The experimental data on kaons and other mesons containing strange quarks are both abundant and rather precise [\refcite{PDG}]. 
These particles are expected to play an important role in meson vacuum phenomenology and at non-zero 
temperatures, most notably in the restoration of the chiral symmetry that is spontaneously [\refcite{Goldstone}] and explicitly [\refcite{Hooft}] broken in vacuum. Their structure - at least in the scalar sector - is ambiguous, as in the case of the non-strange mesons [\refcite{Paper1}].\\
Open questions regarding the strange mesons can be addressed, for example, within the linear sigma model [\refcite{gellmanlevy}], as has been done, e.g., in Ref.\ [\refcite{Reference1}]. However, our approach is different in comparison to Ref.\ [\refcite{Reference1}] in that our calculations are based on a globally chirally invariant $N_f=3$ Lagrangian that contains the scalar and pseudoscalar but also vector and axial-vector degrees of freedom. In this paper, we briefly outline our calculations and present some first results. The paper is organized as follows: in Sec.\ 2 we present the model Lagrangian and its implications and in Sec.\ 3 we summarize our results.

\section{The Model and its Implications}

The Lagrangian of the $N_f=3$ model with global chiral invariance has an analogous form as the corresponding linear sigma model Lagrangian for $N_f=2$ [\refcite{Paper1}]:
\begin{align}
\mathcal{L}  &  =\mathrm{Tr}[(D^{\mu}\Phi)^{\dagger}(D^{\mu}\Phi)]-m_{0}^{2}
\mathrm{Tr}(\Phi^{\dagger}\Phi)-\lambda_{1}[\mathrm{Tr}(\Phi^{\dagger}%
\Phi)]^{2} -\lambda_{2}\mathrm{Tr}(\Phi^{\dagger}\Phi)^{2}\nonumber\\
&  -\frac{1}{4}\mathrm{Tr}[(L^{\mu\nu})^{2}+(R^{\mu\nu})^{2}]+\frac{m_{1}^{2}%
}{2} \mathrm{Tr}[(L^{\mu})^{2}+(R^{\mu})^{2}]+\mathrm{Tr}[H(\Phi+\Phi
^{\dagger})]\nonumber\\
&  +c(\det\Phi+\det\Phi^{\dagger})-2ig_{2}(\mathrm{Tr}\{L_{\mu\nu}[L^{\mu
},L^{\nu}]\} +\mathrm{Tr}\{R_{\mu\nu}[R^{\mu},R^{\nu}]\})\nonumber\\
&  -2g_{3}\left[  \mathrm{Tr}\left(  \left\{  \partial_{\mu}L_{\nu}+\partial_{\nu}L_{\mu} \right\}
\{L^{\mu},L^{\nu}\}\right) + \mathrm{Tr} \left(  \left\{  \partial_{\mu}%
R_{\nu} +\partial_{\nu}R_{\mu} \right\}  \{R^{\mu},R^{\nu}\}\right)  \right]  \nonumber\\
&  +\frac{h_{1}}{2}\mathrm{Tr}(\Phi^{\dagger}\Phi)\mathrm{Tr}[(L^{\mu})^{2}
+(R^{\mu})^{2}]+h_{2}\mathrm{Tr}[(\Phi R^{\mu})^{2}+(L^{\mu}\Phi)^{2}]
+2h_{3}\mathrm{Tr}(\Phi R_{\mu}\Phi^{\dagger}L^{\mu}). \label{Lagrangian}%
\end{align}
where
\begin{equation}
\Phi := \frac{1}{\sqrt{2}}\left(
\begin{array}
[c]{ccc}%
\frac{(\sigma_{N}+a_{0}^{0})+i(\eta_{N}+\pi^{0})}{\sqrt{2}} & a_{0}^{+}%
+i\pi^{+} & K_{S}^{+}+iK^{+}\\
a_{0}^{-}+i\pi^{-} & \frac{(\sigma_{N}-a_{0}^{0})+i(\eta_{N}-\pi^{0})}%
{\sqrt{2}} & K_{S}^{0}+iK^{0}\\
K_{S}^{-}+iK^{-} & {\bar{K}_{S}^{0}}+i{\bar{K}^{0}} & \sigma_{S}+i\eta_{S}%
\end{array}
\right) \label{Phi}
\end{equation}
is a matrix containing the scalar and pseudoscalar degrees of freedom and
\begin{align}
L^{\mu} := \frac{1}{\sqrt{2}}\left(
\begin{array}
[c]{ccc}%
\frac{\omega_{N}^{\mu}+\rho^{\mu0}}{\sqrt{2}}+\frac{f_{1N}^{\mu}+a_{1}^{\mu0}%
}{\sqrt{2}} & \rho^{\mu+}+a_{1}^{\mu+} & K_{\star}^{\mu+}+K_{1}^{\mu+}\\
\rho^{\mu-}+a_{1}^{\mu-} & \frac{\omega_{N}^{\mu}-\rho^{\mu0}}{\sqrt{2}}%
+\frac{f_{1N}^{\mu}-a_{1}^{\mu0}}{\sqrt{2}} & K_{\star}^{\mu0}+K_{1}^{\mu0}\\
K_{\star}^{\mu-}+K_{1}^{\mu-} & {\bar{K}}_{\star}^{\mu0}+{\bar{K}}_{1}^{\mu0}
& \omega_{S}^{\mu}+f_{1S}^{\mu}%
\end{array}
\right), \nonumber \\
R^{\mu} := \frac{1}{\sqrt{2}}\left(
\begin{array}
[c]{ccc}%
\frac{\omega_{N}^{\mu}+\rho^{\mu0}}{\sqrt{2}}-\frac{f_{1N}^{\mu}+a_{1}^{\mu0}%
}{\sqrt{2}} & \rho^{\mu+}-a_{1}^{\mu+} & K_{\star}^{\mu+}-K_{1}^{\mu+}\\
\rho^{\mu-}-a_{1}^{\mu-} & \frac{\omega_{N}^{\mu}-\rho^{\mu0}}{\sqrt{2}}%
-\frac{f_{1N}^{\mu}-a_{1}^{\mu0}}{\sqrt{2}} & K_{\star}^{\mu0}-K_{1}^{\mu0}\\
K_{\star}^{\mu-}-K_{1}^{\mu-} & {\bar{K}}_{\star}^{\mu0}-{\bar{K}}_{1}^{\mu0}
& \omega_{S}^{\mu}-f_{1S}^{\mu}%
\end{array}
\right) \label{LR}
\end{align}
are, respectively, the left-handed and right-handed matrices containing the vector and axial-vector degrees of freedom.
Furthermore, $D^{\mu}\Phi=\partial^{\mu}\Phi-ig_{1}(L^{\mu}\Phi-\Phi R^{\mu})$ is the covariant derivative; $L^{\mu\nu} = \partial^{\mu}L^{\nu} - \partial^{\nu}L^{\mu}$ and $R^{\mu\nu} = \partial^{\mu}R^{\nu} - \partial^{\nu}R^{\mu}$ are, respectively, the left-handed and right-handed field strength tensors; the term Tr$[H(\Phi+\Phi^{\dagger})]$ [$H = 1/2 \, {\rm diag}(h_{0N},h_{0N},\sqrt{2} h_{0S})$, $h_{0N}=const.$, $h_{0S}=const.$] explicitly breaks chiral symmetry due to nonzero quark masses, and the term $c\,(\det\Phi+\det\Phi^{\dagger})$ describes the $U(1)_A$ anomaly [\refcite{Hooft}].\\
As in Ref.\ [\refcite{Paper1}], in the non-strange sector, we assign the fields $\vec{\pi}$
and $\eta_{N}$ to the pion and the $SU(2)$ counterpart of the
$\eta$ meson, $\eta_{N}\equiv(\overline{u}u+\overline{d}d)/\sqrt{2}$. The fields $\omega^{\mu}_N$ and $\vec{\rho}^{\mu}$ represent the
$\omega(782)$ and $\rho(770)$ vector mesons, respectively, and the fields
$f_{1 N}^{\mu}$ and $\vec{a}_{1}^{\mu}$ represent the $f_{1}(1285)$ and
$a_{1}(1260)$ mesons, respectively. In the strange sector, we assign the $K$ fields to the kaons; the $\eta_S$ field is the strange contribution to the physical $\eta$ and $\eta '$ fields; the $\omega_S$, $f_{1S}$, $K_{\star}$, and $K_1$ fields correspond to the $\phi(1020)$, $f_1(1420)$, $K^{\star}(892)$, and $K_1(1270)$ mesons, respectively. In accordance with Ref.\ [\refcite{Paper1}], we assign the scalar kaon $K_S$ to the physical $K_0^{\star}(1430)$ state. The assignment of the non-strange scalar states depends on the results for their masses (see next paragraph).\\
In order to implement spontaneous symmetry breaking, we shift the $\sigma_N$ and $\sigma_S$ fields by their respective vacuum expectation values $\phi_N$ and $\phi_S$, resulting in $\eta_N$-$f_{1N}$ and $\vec \pi$-$\vec a_1$ mixing [\refcite{Paper1}] as well as in $\eta_S$-$f_{1S}$, $K_S$-$K_{\star}$, and $K$-$K_1$ mixing. These are removed as described in Ref.\ [\refcite{Paper1}]. We observe consequently the mixing between the $\sigma_N$ ($\equiv n \bar{n}$, with $n$ standing for non-strange quarks) and $\sigma_S$ ($=s \bar{s}$) states, yielding two scalar states that are respectively predominantly of $n \bar{n}$ or $s \bar{s}$ nature [note that the Lagrangian (\ref{Lagrangian}) also yields $\eta_N$ - $\eta_S$ mixing]. In this paper, it suffices to say that our preliminary results, in the limit where the large-$N_C$ suppressed parameter $\lambda_1$ is set to zero, show the existence of a predominantly $n \bar{n}$ state with a mass between 1.0 and 1.3 GeV and of a predominantly $s \bar{s}$ state with mass between 1.3 and 1.6 GeV, depending on the choice of parameters. Our work concerning these (and other) issues is still ongoing and will allow us to determine the parameters in the model more exactly so that the masses of the two mentioned scalars will become more precisely determined [\refcite{PGRKW}]. Note that our $N_f=3$ model possesses no free parameters due to the preciseness of the experimental data [\refcite{PDG}].

\section{Summary and Outlook}

We have presented an $N_f=3$ linear sigma model with (axial-)vector degrees of freedom and global chiral invariance. Our preliminary results suggest that the notion of the scalar states above 1 GeV as quarkonia may hold also when the $N_f=2$ Lagrangian of Ref.\ [\refcite{Paper1}] is extended to three flavors. Clearly, the parameters in our model still need to be calculated precisely and, regarding the scalar sector, one needs to include the tetraquark and glueball fields in order to study their mixing with the scalar quarkonia.

\section{References}

\end{document}